\documentclass[twocolumn, twocolappendix]{aastex7}
\usepackage{xcolor}
\definecolor{blazeorange}{rgb}{1.0, 0.4, 0.0}
\definecolor{seagreen}{rgb}{0.18, 0.55, 0.34}
\definecolor{rufous}{rgb}{0.66, 0.11, 0.03}
\definecolor{royalfuchsia}{rgb}{0.79, 0.17, 0.57}
\definecolor{scarlet}{rgb}{1.0, 0.13, 0.0}
\definecolor{royalpurple}{rgb}{0.47, 0.32, 0.66}
\definecolor{darkblue}{rgb}{0, 0, 0.66}

\usepackage{graphicx}	
\usepackage{amsmath}	
\usepackage{amssymb}	
\usepackage{xcolor}
\usepackage{ulem}
\usepackage{hyperref}

\begin{document}

\title{The Role of Scintillation in Detecting HI Absorption in FRB Spectra}

\correspondingauthor{Om Gupta}

\author[0000-0001-8470-7289]{Om Gupta}
\affiliation{Department of Astronomy, The University of Texas at Austin, Austin, TX 78712, USA}
\email[show]{omgupta@utexas.edu}

\author[0009-0003-8955-7402]{Pawan Kumar}
\affiliation{Department of Astronomy, The University of Texas at Austin, Austin, TX 78712, USA}
\email{pk@astro.as.utexas.edu}

\author[0000-0001-7833-1043]{Paz Beniamini}
\affiliation{Department of Natural Sciences, The Open University of Israel, PO Box 808, Ra’anana 4353701, Israel}
\affiliation{Astrophysics Research Center of the Open University (ARCO), The Open University of Israel, PO Box 808, Ra’anana 4353701, Israel}
\affiliation{Department of Physics, The George Washington University, 725 21st Street NW, Washington, DC 20052, USA}
\email{pazb@openu.ac.il}

\author[]{Edward L. Robinson}
\affiliation{Department of Astronomy, The University of Texas at Austin, Austin, TX 78712, USA}
\email{rob@astro.as.utexas.edu}

\begin{abstract}
The 21 cm absorption line of neutral hydrogen has been a long hypothesized observational feature of fast radio burst (FRB) spectra. Difficulties associated with detector noise in extracting HI absorption have been previously studied. We test the role that scintillation plays in its detectability and characterize the regimes where a realistic FRB may yield the 21 cm line. We build an efficient model to simulate diffractive scintillation arising from FRB passage through a thin scattering screen. We find that the absorption profile is detectable in a scintillation-dominated high signal-to-noise spectrum if the scintillation decorrelation bandwidth differs significantly in scale from absorption profile width. Specifically, for a detector with a known SEFD, the HI line can be extracted using a low-pass filter for scintillation bandwidths $< 0.5$ kHz, peak cloud optical depths $> 0.5$, and fluences $\gtrsim$ SEFD Jy ms (assuming a 1 ms duration). Stacking repeat bursts from active repeaters is also useful, but they must be separated in time by more than the diffractive scintillation timescale, otherwise flux modulations with frequency are correlated. By cross-referencing repeating FRB positions with a catalog of Milky Way molecular clouds detected in CO, we find that the sightline to FRB 20180916B may intersect a Galactic molecular cloud. This possibility may be confirmed with a hundred or more baseband-recorded bursts with GBT. We also identify clear observation strategies with FAST and GBT for a range of repeater luminosity functions, with both telescopes needing roughly 500 bursts to confidently detect a peak optical depth of $0.5$.
\end{abstract}



\section{Introduction} \label{sec:intro}
Fast radio bursts (FRBs) are highly energetic radio signals that are typically extragalactic in origin. They are detected ubiquitously across the sky and have been observed between 110 MHz - 8 GHz. The 21 cm (or 1.4 GHz) hyperfine transition of neutral hydrogen is one of the most well-known features of radio astronomy, which is seen both in emission as well as in absorption against background radio sources. 

Because FRBs are detected in the radio-band, the signals may pass through structures containing neutral hydrogen, located somewhere along the sightline. Whether HI is observed in emission or absorption depends on the background radio signal. Since typical brightness temperatures for FRBs are $10^{36}$ K, HI hyperfine transitions will leave an absorption imprint on the FRB spectrum.

There are a large number of molecular clouds, primarily in the disk of the Milky Way, which host cold HI gas along with molecular hydrogen \citep{Dame2001}. Similarly, FRB host galaxies exhibit diverse properties in their HI content and distribution (\citealt{Glowacki2023}, and references therein), as well as in their molecular gas content, as traced by CO \citep{Hatsukade2022, Chittidi2023, Hsu2023}. Although, the focus of these studies is on the properties and kinematics of HI and H$_2$ in the whole of the galaxy.

Pulsars have been used to detect HI absorption for decades (e.g., \citealt{Ables1976, Weisberg1980, Weisberg2008, Jing2023}). The possibility of FRBs being used to detect HI absorption imprinted on their spectra was first discussed by \citet{Fender2015}. Like that study, our primary focus will be on cold HI gas in the Milky Way, the host galaxy and intervening galaxies, as the source of HI absorption. Detectable HI absorption in FRB host galaxies can directly yield the redshift of the host without the need for localization \citep{Margalit2016}. However, with significantly high localization precision and optical follow-up studies, obtaining redshifts for host galaxies nowadays is relatively easier, for both repeaters and nonrepeaters. 

FRBs can serve as alternate probes of neutral hydrogen in our own galaxy for low-latitude sightlines, which effectively intersect one or more molecular cloud regions in the Milky Way disk. But the more unique opportunity is that the pencil-beam FRB sightline can probe the chemical features of individual cold gas clouds in different galaxies on a subcloud transverse scale. Even though, as we show later, the fraction of random sightlines intersecting cold HI clouds in a galaxy is low, usable cases can be found within the rapidly growing FRB population. 
Furthermore, it may prove useful for high-redshift applications in the future, such as in detecting a high-redshift forest of HI absorption lines arising from neutral hydrogen structures called minihalos \citep{Furlanetto2002}. With 21 cm detection from minihalos, constraining the thermal history of the early Universe, specifically during the Epoch of Reionization, would become possible.

\subsection{Neutral Hydrogen Absorption}
HI is an abundant species in molecular cloud halos, and is observed with large optical depths in surveys (e.g., \citealt{Rugel2018}), which find neutral hydrogen column densities upwards of $10^{21} \rm cm^{-2}$. The hydrogen column densities are typically defined in these surveys as:
\begin{equation}
    N_{\rm HI} = 1.82\times10^{18} \frac{T_{\rm spin}}{1 \;\rm K} \int \tau(v) \frac{dv}{1\;\rm km \, s^{-1}} {\rm cm^{-2}} \label{eq:HI_coldepth}
\end{equation}
Here, $T_{\rm spin}$ is the spin temperature of the gas, assumed to be constant throughout the molecular cloud, and $\tau(v)$ is the optical depth as a function of line-of-sight velocity $v$ (or equivalently as a function of frequency).
There are certain caveats to this estimate. Although Equation (\ref{eq:HI_coldepth}) is formally general, in practice we measure the brightness temperature rather than the optical depth directly.
As a result, applying this expression requires the optically thin approximation. However, this assumption is often violated, as optical depths $\tau \gtrsim 1$ are commonly observed in such clouds. Consequently, the inferred column densities should be regarded as lower limits.

Alternatively, we can use the hyperfine level-splitting calculations to define the optical depth at line center:
\begin{equation}
    \tau_0 = 0.31 \Big[\frac{N_{\rm HI}}{10^{21} \,{\rm cm^{-2}}} \Big] \Big[\frac{10 \,{\rm km s^{-1}}}{b} \Big] \Big[\frac{10^2 \,{\rm K}}{T_{\rm spin}} \Big],
    \label{eq:tau0}
\end{equation}
where, $b$ is the velocity broadening of the optical depth profile.
The broadening seen in molecular clouds is often much larger than the calculated thermal broadening and is attributed to turbulent motions within the cloud. \cite{Heyer2004} found that
\begin{equation}
    b = \sqrt{2}\sigma_v = 0.96 \, {\rm km \, s^{-1}} \Big(\frac{d}{1 \, {\rm pc}} \Big)^{0.59}
    \label{eq:velocity_dispersion}
\end{equation}
where $d$ is the size of the molecular cloud.
The frequency dependence of the absorption is given by the Voigt profile (see e.g., \citealt{Ryden2021_book}), and the full width half-maximum (FWHM) of the absorption profile $\approx 2\sqrt{\ln 2} (b/c)\nu_0 \approx (7.57 \;{\rm kHz}) (d/1\,{\rm pc})^{0.65}$.

\section{Scintillation}
As FRBs pass through the ionized material in the ISM of the Milky Way, they are scattered due to the changes in the density of gas clumps along their path. The wave front gets distorted as the electric field vectors at different points accumulate different phase shifts. Furthermore, this phase shifting is frequency dependent, due to which the electric field vectors adding up to constitute the signal at the observer's location within the time resolution of the detector create a scintillation pattern in which the flux varies with frequency.

Scintillation is a common phenomenon observed in the radio for Galactic point sources such as pulsars, and for extragalactic point sources like FRBs \citep{Narayan1992, Cordes2019}. It is also a very powerful tool that can tell us about FRB physics \citep{Beniamini2020,Kumar2024, Nimmo2025}, the host galaxy and Milky Way ISM properties, and the source-screen-observer geometries \citep{Sammons2023, Pradeep2025, Scott2025}. 
Even though scintillation arises from the inhomogeneity of the electron distribution and its temporal change in a 3D volume, its properties can be typically encapsulated by modeling a thin scattering screen. 
For FRBs, the electrons that contribute to scintillation/scattering are usually mostly concentrated in the Milky Way and the host galaxy, for which separate scattering screens must be invoked to explain scintillation behavior (although some cases can be explained with just one screen).

\subsection{Can Scintillation Destroy Absorption Features in FRB Spectra?} \label{sec:scint}
The variation of flux with frequency is correlated over a certain frequency scale called the scintillation decorrelation bandwidth $\delta \nu_{\rm dc}$ \citep{Narayan1992}. Scintillation bandwidth in the Milky Way generally follows a power law of the form $\delta \nu_{\rm dc} \propto \nu^{\beta}$, with $\beta = 4.4$ for a Kolmogorov distribution. However, deviations from this power-law index are known to exist \citep{Schoen2021}. For example, FRB 20201124A has $\beta=3.5\,\pm\,0.1$ from a scattering screen located roughly 400 pc from the Earth \citep{Main2022}. 

If the absorption signature from a cold HI cloud has characteristic width on the order of the scintillation bandwidth, then scintillation caused by turbulence in the screen will introduce large variations in the signal that can wipe out the absorption profiles or make them extremely difficult to detect. We hypothesize, however, that if the absorption profiles are either significantly wider or narrower than the scintillation bandwidth, it is possible to detect an absorption trough. This is subject to the condition that if the absorption features are narrower, then they should not be narrower than the frequency resolution of the detector. The objective of this paper is to show this via simulations and provide a clear pathway for HI absorption line detection in FRB spectra. We next outline a method to simulate a thin scattering screen and explore the regimes and effects of scintillation on absorption line detection. The results are discussed in Section \ref{sec:recover_line} in the context of the number and placement of observationally relevant screens.

\section{Modeling scintillation from a thin scattering screen} \label{sec:scint_computation}
An important lengthscale in modeling a scintillating screen is the Fresnel scale,
\begin{equation}
    R_F = \sqrt{\frac{\lambda d_{\rm so} d_{\rm fs}}{d_{\rm fo}} },
    \label{eq:fresnel_scale}
\end{equation}
which is the transverse lengthscale on the screen that marks the transition from strong to weak diffractive scintillation. It differentiates between regimes where flux modulation is strong to where it is weak. In Equation (\ref{eq:fresnel_scale}), $d_{\rm fo}$ refers to the FRB-observer distance, $d_{\rm so}$ to the screen-observer distance, and $d_{\rm fs}$ to the FRB-screen distance. 

For a thin scattering screen, it is assumed that the phase shift is contributed by turbulent eddies within the screen, such that the phase difference between two points, located on the face of the screen toward the observer, increases with transverse distance between those points. 
We define the diffractive scale $\ell_\pi$ as the transverse scale on the screen upto which the phase fluctuations from points are coherent to within $\pi$ radians. This is denoted as $\Delta \phi_s(\ell_\pi) \sim \pi$, where $\phi_s$ represents the phase shift after crossing the screen.
For Kolmogorov density fluctuations, $\ell_{\pi} \propto \lambda^{-6/5}$ \citep{Beniamini2020}. The simulation is designed for strong diffractive scintillation whose conditions are that $\ell_{\pi} \ll R_F$ and the source size is smaller than $\ell_\pi$. The screen properties are set in such a way that the first condition is satisfied. FRBs are point sources for any screen in the Milky Way, and we assume them to be point sources if a host galaxy screen is being simulated, based on \citet{Nimmo2025}, for the second condition to be satisfied.

The thin screen assumption is valid in most cases for FRB scattering in the interstellar medium, because the screen thickness of $\sim {\rm few} \times {\rm kpc}$ is much smaller than typical path lengths $d_{\rm fo} \sim$ Gpc. The effective distance in Equation (\ref{eq:fresnel_scale}) is hence either $\sim d_{\rm so}$ or $\sim d_{\rm fs}$. As a single ray traverses this screen, it passes through a large number of turbulent eddies, which act as the density perturbations that scatter the ray. Thus, the phase accumulated by a ray emerging from a particular point on the screen is a Gaussian random variable. Choosing a reference point aligned with the central sightline of the FRB, we define the turbulent scattering contributed phase at this point to be $\phi_{\rm s}(r=0)$. The phase at a transverse distance $r$ is defined to be $\phi_{\rm s}(r)$, and, $\Delta \phi_{\rm s}(r) \equiv \phi_{\rm s}(r) - \phi_{\rm s}(r=0)$ then defines a Gaussian random field. To construct a distribution for phase differences on the screen, we utilize the two-point correlation function, in the form of the phase structure function $D_{\phi}(r)$, which is defined as,
\begin{equation}
    D_{\phi}(r) = \langle [\phi_{\rm s}(r) - \phi_{\rm s}(r=0)]^2 \rangle = \pi \Big( \frac{r}{\ell_{\pi}} \Big)^{5/3}.
\end{equation}
The last equality in the above equation arises from an assumption of isotropic Kolmogorov turbulence \citep{Narayan1992}. For an alternative model, like the square-law model for which $\beta=4$, $D_{\phi} \propto r^2$ \citep{Cordes1998, Lambert2000}. The distribution of phases contributed by the scattering screen, for each separation $r$, is a normal distribution with zero mean and standard deviation equal to $\sqrt{D_{\phi}(r)}$. Different frequencies pass through the turbulent eddies within a timespan shorter than it takes for the eddies to change, on average, causing the scattering phase to be related across frequencies as $\phi_{\rm s}(\nu_1) = \phi_{\rm s}(\nu_0) \nu_0 / \nu_1$. This frequency dependence arises from the definition of the rms phase shift after traversing the thickness of the scattering screen \citep{Beniamini2020} and is valid for $|\nu_1-\nu_0|\ll\nu_0$.

In addition to the scattering phase, there is a geometric phase associated with the extra distance scattered light rays have to travel. This is given by \citep{Narayan1992},
\begin{equation}
    \phi_{\rm g} = \pi \Big( \frac{r}{R_F} \Big)^2.
    \label{eq:geom_phase}
\end{equation}
The size of the screen that is visible to the observer is, also known as the refractive scale,
\begin{equation}
    R_{\rm scat} = \frac{R_F^2}{\pi \ell_{\pi}}.
    \label{eq:screen_size}
\end{equation}
The scintillation decorrelation bandwidth is estimated as,
\begin{equation}
    \delta \nu_{\rm dc} = \pi \nu \Big( \frac{\ell_\pi}{R_F} \Big)^2.
    \label{eq:decorr_bandwidth}
\end{equation}
The above Equations (\ref{eq:screen_size}) and (\ref{eq:decorr_bandwidth}) have been taken from \citet{Beniamini2020}.

\begin{figure}
    \centering
    \includegraphics[width=0.5\linewidth]{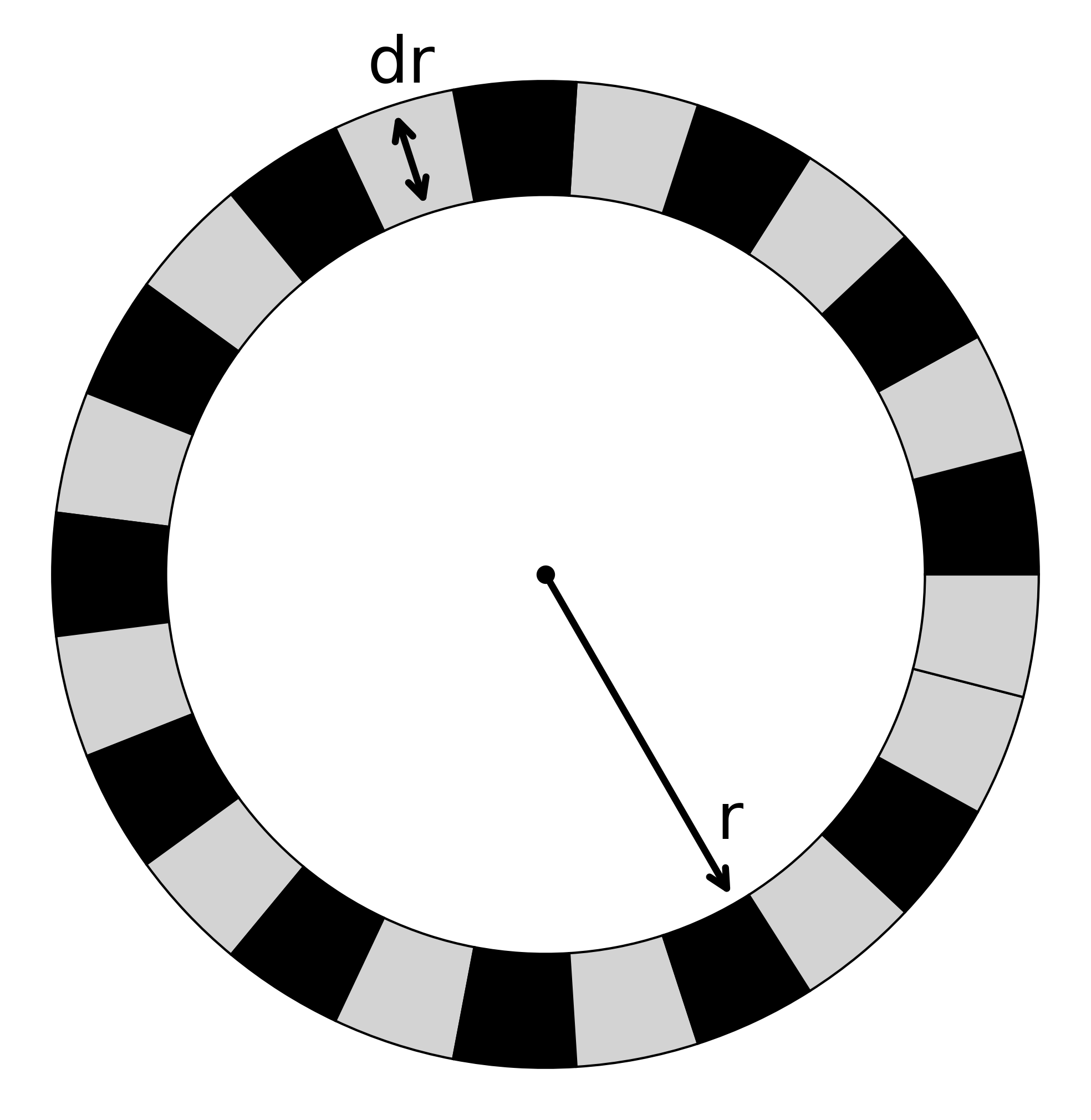}
    \caption{The schematic depicts a ring on the scattering screen at a radius $r$ and thickness $dr\;(=1)$. As an example, it has been divided into 25 patches, which corresponds to $r=25/2\pi\approx 4$. The number of patches used to calculate the Fresnel Kirchoff integral is $n_{\rm polar} \approx 12$, which are colored black.}
    \label{fig:ring}
\end{figure}

To simulate the scintillation pattern, scale free coordinates are chosen, and the scale of the system $l_0$ is included to compute physical distances, if required. Employing polar coordinates, the plane of the screen is divided into rings of thickness $dr$ at a distance $r$ from the center. Each ring is further divided up into $2\pi r/dr$ chunks of area $\approx dr^2$, but only $n_{\rm polar}$ chunks are used, where $n_{\rm polar} = 2\pi \sqrt{r} / dr$. See Figure \ref{fig:ring} for an illustrative example.

Using $\sqrt{r}$ instead of $r$ reduces the number of computations to be performed as $r$ becomes large. This does not bias the results as sampling less points at each $r$ still yields the underlying distribution, that is, a Gaussian with zero mean and $D_{\phi}$ standard deviation\footnote{We are sampling independent phase structure Gaussians for each ring. Therefore, two spatially close patches on nearby rings will not necessarily exhibit phase correlation, even though they should, with the phase being a Gaussian random field. However, in the end, patches on a ring are not assigned a unique position on the screen, which bypasses the requirement of phase correlation between two patches, as long as neither of the two is the central patch.}. Sampling more points is preferable to get closer to the phase structure Gaussian, but we deem this approach to be an acceptable tradeoff for the large computation time otherwise required.

Using the form of the Fresnel-Kirchhoff integral defined in \citet{Born1959} and \citet{Narayan1992}, we have,
\begin{equation}
    \psi(r, \nu) \propto \frac{1}{2\pi R_F^2} \sum_{i=1}^{n_{\rm polar}} \exp [\iota(\phi_{\rm s,i} + \phi_{\rm g})] dr^2,
    \label{eq:fresnel_kirchoff_integral}
\end{equation}
where, $dr^2$ is the differential area element on the face of the screen. When simulating scintillation from many bursts, some further steps are implemented to reduce the computation time that have negligible effects on the result. These steps are outlined in Appendix \ref{sec:reduce_comp}, and help to decrease the computational requirement by a factor of more than $10^4$.
The total amplitude of the electric field at the observer is $\Psi(\nu) = \sum_r \psi(r,\nu)$, and the specific intensity $I(\nu) = |\Psi(\nu)|^2$. The autocorrelation function $A(\Delta \nu) = \langle I(\nu) I(\nu+\Delta \nu) \rangle$ is calculated for a range of $\Delta \nu$ values. Following the methodology of how the scintillation bandwidth is calculated for real data, the simulated spectrum is fit with a Lorentzian, given by,
\begin{equation}
    \mathcal{L}(\Delta \nu) = \frac{m_I}{1+( \Delta \nu/\delta \nu_{\rm sdc})^2}.
    \label{eq:lorentzian_acf}
\end{equation}
Here, $m_I \in [0,1]$ is the modulation index that indicates the level of flux modulation and $\delta \nu_{\rm sdc}$ is the decorrelation bandwidth recovered from the simulated data. We note here that Kolmogorov turbulence does not exactly yield a Lorentzian ACF, but it is close to being so. The recovered scintillation bandwidths using Equation (\ref{eq:lorentzian_acf}) are seen to be close to the input value on average, although they can vary by $\sim 10\%$. Furthermore, we also compare the modulation indices and scintillation bandwidths recovered for a sample of 100 bursts generated using $n_{\rm polar}=2\pi r /dr$ and using $n_{\rm polar} = 2\pi \sqrt{r} / dr$. The latter is found to be acceptably close to the former.

A recent work used a publicly available code, called \textsc{ScintillationMaker}\footnote{https://github.com/SprengerT/ScintillationMaker/tree/main}, to synthesize the complex electric field result from diffractive scintillation \citep{Balzan2026}. \textsc{ScintillationMaker} uses the stationary-phase approximation to calculate the resultant electric field from uniform randomly distributed images on the scattering screen. The amplitude of contribution of each image to the total sum that is the electric field is sampled from a Gaussian distribution, with images further away on the screen from the central sightline having a smaller amplitude. We verify all results with this tool, to ensure robustness of our conclusions irrespective of the screen simulation method used.
Readers are also directed to \citet{Hamidouche2007} for another simulation methodology that applies the Kolmogorov spectrum on spatial Fourier transforms. 

The timescale over which the scintillation pattern changes is called the scintillation timescale. Our simulation methodology is limited to only produce independent uncorrelated strong diffractive scintillation snapshots. This works because millisecond FRB durations are effectively instantaneous compared to diffractive and refractive scintillation timescales. However, \textsc{Scintillation Maker} can handle the diffractive scintillation timescale, and the methodology of \citet{Hamidouche2007} can additionally handle refractive scintillation and its timescale.

\section{Recovering HI Absorption Lines in Simulated Scintillation} \label{sec:recover_line}
\cite{Cordes2019} show, using the NE2001 model \citep{Cordes2002}, that Galactic scintillation bandwidth depends strongly on both latitude and longitude and decreases significantly from $\sim$MHz range to sub-KHz range from high to low latitudes. To detect HI absorption, the range of Galactic latitudes and longitudes should be in a regime where Galactic screen scintillation bandwidths have a significantly different scale than the HI absorption profile widths. 

In the Milky Way, and also in typical galaxies, cold HI regions/clouds are mostly distributed at low Galactic latitudes. For example, 88.5\% of sightlines in the sky area surveyed by \citet{Dame2001} within Galactic latitudes $-5^{\circ} < b < 5^{\circ}$ pass through molecular clouds \citep{Miville2017}. Using a simple cloud distribution model, described in Appendix \ref{app:galdist_model}, it is shown that the average number of molecular clouds intercepted by a sightline is $\geq 1$ at $b \lesssim 2^{\circ}$. As such, an FRB may have an HI absorption line imprinted on its spectrum, if its sightline is seen either at a low latitude in the Milky Way, or passes through another galaxy oriented at a low inclination angle. Absorption external to the Milky Way can either occur in the host galaxy itself, or in an intervening galaxy.

Generally, HI absorption profiles are a few tens of kHz wide, and consequently, Galactic clouds would be best observed at low latitudes with sub-kHz scintillation bandwidths. For example, sightlines exhibit $\delta \nu_{\rm dc} \approx 0.01 - 10$ kHz at $b=1^{\circ}$ and $1.42$ GHz, with larger longitudes having a higher $\delta \nu_{\rm dc}$ (estimated using NE2025; \citealt{Ocker2026}).
In this case, scintillation from the host galaxy should also not have the same scale as the absorption profile. To detect absorption from a known host or intervening galaxy that is at a low inclination angle, the best case scenario is for sightlines out of the plane of the Milky Way, wherein Galactic scintillation will not compete with host/intervening galaxy's HI absorption.

\begin{figure*}
    \centering
    \includegraphics[width=0.49\linewidth]{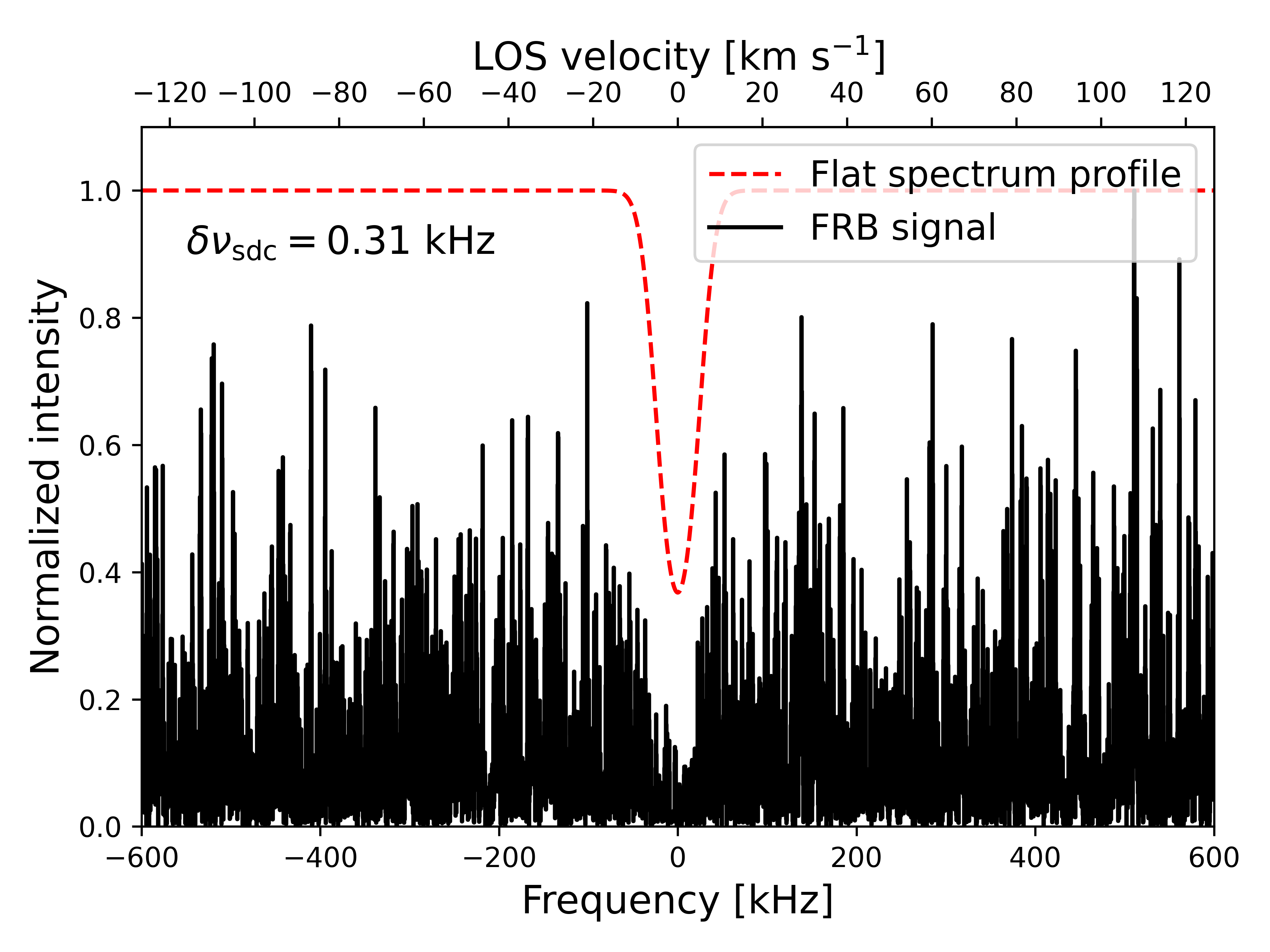}
    \hfill
    \includegraphics[width=0.49\linewidth]{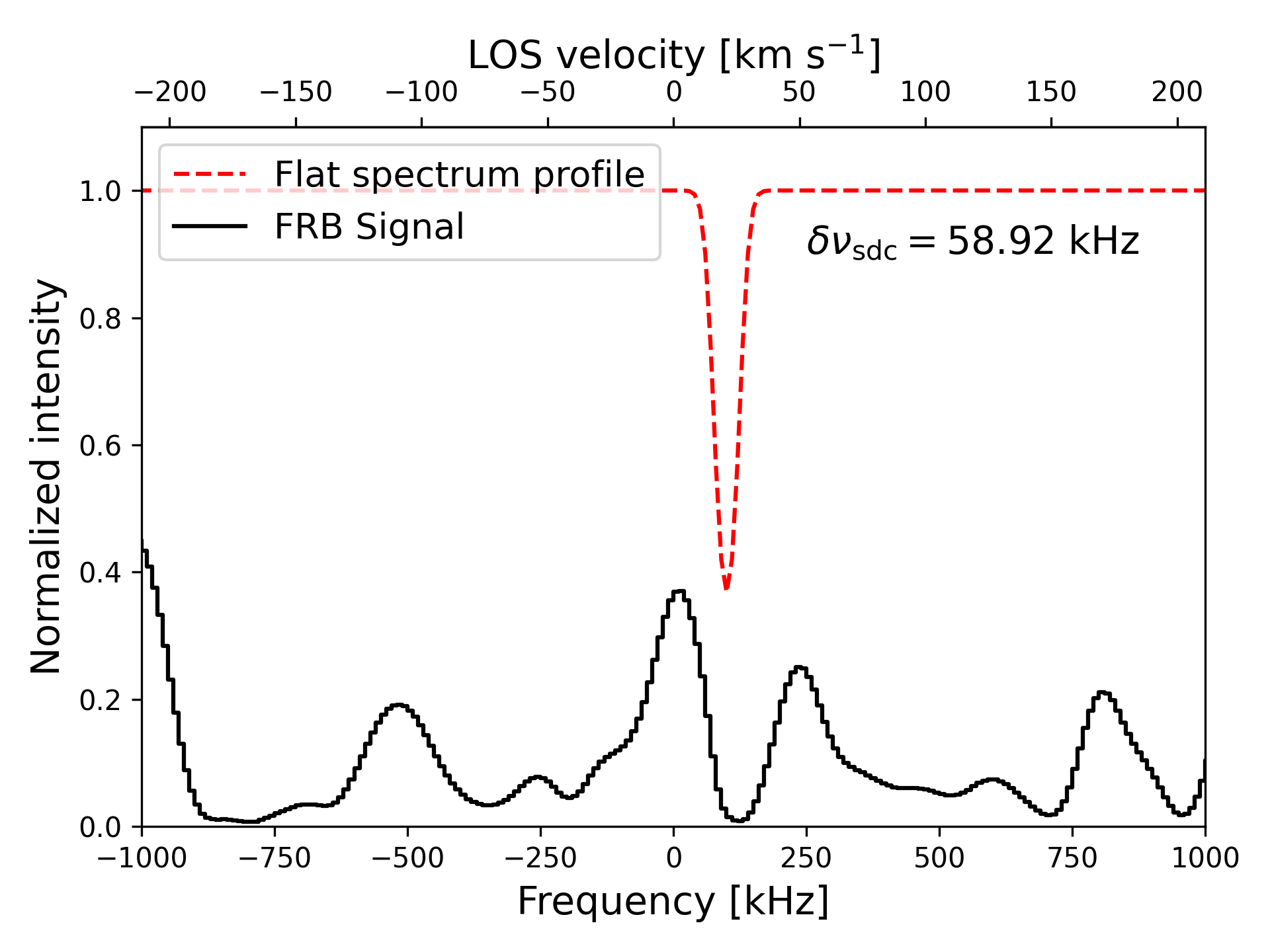}
    \caption{These plots show how FRB spectra modulated by scintillation and without detector noise would look after passing through a cold HI cloud of diameter $d = 20$ pc, $T_{\rm spin} = 50$ K, and $N_{\rm HI} \approx 9 \times 10^{20}$ cm$^{-2}$, yielding $\tau_0 = 1$ and ${\rm FWHM}\approx 44$ kHz. (\textit{left}) The scintillation decorrelation bandwidth $\delta \nu_{\rm sdc} = 0.31$ kHz, being much smaller than the FWHM of the HI profile, allows HI absorption to be distinguishable in the spectrum. The observational frequency resolution $\delta \nu_{\rm obs} = 0.3$ kHz. (\textit{right}) $\delta \nu_{\rm sdc} = 58.92$ kHz, which is of the same order as the FWHM. The central frequency of absorption has been shifted to the right by 100 kHz to emphasize that the profile completely vanishes and is indistinguishable among scintillation. $\delta \nu_{\rm obs} = 10$ kHz.}
    \label{fig:full_abs_prof}
\end{figure*}

A comparison of HI absorption detectability in an intrinsically flat FRB spectrum, between the cases where $\delta \nu_{\rm dc} \ll$ FWHM and where $\delta \nu_{\rm dc} \sim$ FWHM of the absorption profile, is shown in Figure \ref{fig:full_abs_prof}. These plots are generated using the computationally intensive method in Section \ref{sec:scint_computation} and not with the simplification listed in Appendix \ref{sec:reduce_comp}. It is tested that their recovered modulation indices are $\approx 1$, and $\delta \nu_{\rm sdc}$ values are close to the input $\delta \nu_{\rm dc}$ values.

\subsection{Repeating Fast Radio Bursts} \label{sec:repeat_bursts}
What happens when many bursts from a repeating source are stacked? It is easy to study this analytically, by making certain assumptions that are not made in the simulation. Using Equation (\ref{eq:fresnel_kirchoff_integral}), let the electric field at the observer, $\Psi(\nu) = \sum_{j=1}^k A_j [\cos \phi_j + \iota \sin \phi_j] \equiv X + \iota Y$. Because of strong scattering in the screen, the phase fluctuations across the screen are very large, and we can assume that the total phase $\phi_j$ is sampled from a uniform distribution. Then, the expectation values $\langle X \rangle = \langle Y \rangle = 0$. Under the strong scattering regime, we can also assume that stationary-phase points dominate the phase, leading to the phases being independent because the distance between the stationary-phase points is much larger than $\ell_\pi$. Assuming a large value of $k$, the central limit theorem yields variances as $\langle X^2 \rangle = \langle Y^2 \rangle = (\sum_j A_j^2)/2 \equiv \sigma_\Psi^2$. Thus $X, Y \sim \mathcal{N}(0, \sigma_\Psi^2)$, that is, $X$ and $Y$ are random variables of a normal distribution with zero mean and $\sigma_\Psi$ standard deviation. Now we can define a new variable $Z = (X/\sigma_\Psi)^2 + (Y/\sigma_\Psi)^2$, which by construction becomes a random variable of the chi-squared distribution with 2 degrees of freedom ($\chi_2^2$). Knowing its properties, the observed intensity $I(\nu) = \sigma_\Psi^2 Z$ with $\langle I \rangle=2 \sigma_\Psi^2$ and $\langle I^2 \rangle=4 \sigma_\Psi^4$. The total intensity after stacking $N$ FRBs is $S= \sum_{j=1}^N I_j = \sigma_\Psi^2 \sum_j Z_j$, with $S$ being sampled from the distribution $\sigma_\Psi^2 \chi_{2N}^2$, where $2N$ represents the degrees of the chi-squared distribution. The expected average intensity $\langle \bar{I} \rangle = \langle S \rangle /N = 2\sigma_\Psi^2$, while the standard deviation is $\sigma_{\rm stack} = \langle \bar{I}^2 \rangle^{1/2} = \langle S^2 \rangle^{1/2} / N = 2 \sigma_\Psi^2/N^{1/2}$. The modulation index of the stack is computed as
\begin{equation}
    m_{I, \rm stack} = \frac{\sigma_{\rm stack}}{\langle \bar{I} \rangle} = \frac{1}{\sqrt{N}}.
    \label{eq:modulation_index_stack}
\end{equation}
With more bursts being stacked, the flux modulations decrease. The result is valid only for an ensemble of bursts with equal intrinsic brightness and is independent of the decorrelation bandwidth. If, for example, the flux distribution is a power law with a negative index, as is common for FRBs, then few bright bursts dominate the frequency-dependent fluctuations. This increases the number of bursts required to decrease the modulations by the same amount (quantitatively shown in Section \ref{sec:repeat_bursts}). Using a square-law structure function ($D_\phi \propto r^2$) has no effect on the intensity statistics. With the intensity still being a chi-squared random variable, Equation (\ref{eq:modulation_index_stack}) holds for the square-law model as well.

As noted before, the diffractive scintillation pattern for bursts is correlated within a time period called the diffractive scintillation timescale. This timescale is dependent on the transverse speed of turbulent eddies in the scattering screen. It is typically on the order of minutes for typical screen parameters \citep{Beniamini2020} and has been estimated to be 14.3 min at 1370 MHz for the active repeater FRB 20201124A \citep{Main2022}. Similarly, correlations in closely spaced bursts have been seen in FRB 2020120E \citep{Nimmo2022}. In order to extract a good signal-to-noise HI absorption, those bursts that are observed at time separations greater than the diffractive scintillation timescale must be analyzed. This step is crucial, especially for hyperactive repeaters; otherwise, correlated structures can be spuriously imprinted on the stacked spectrum.

\begin{figure}
    \centering
    \includegraphics[width=\linewidth]{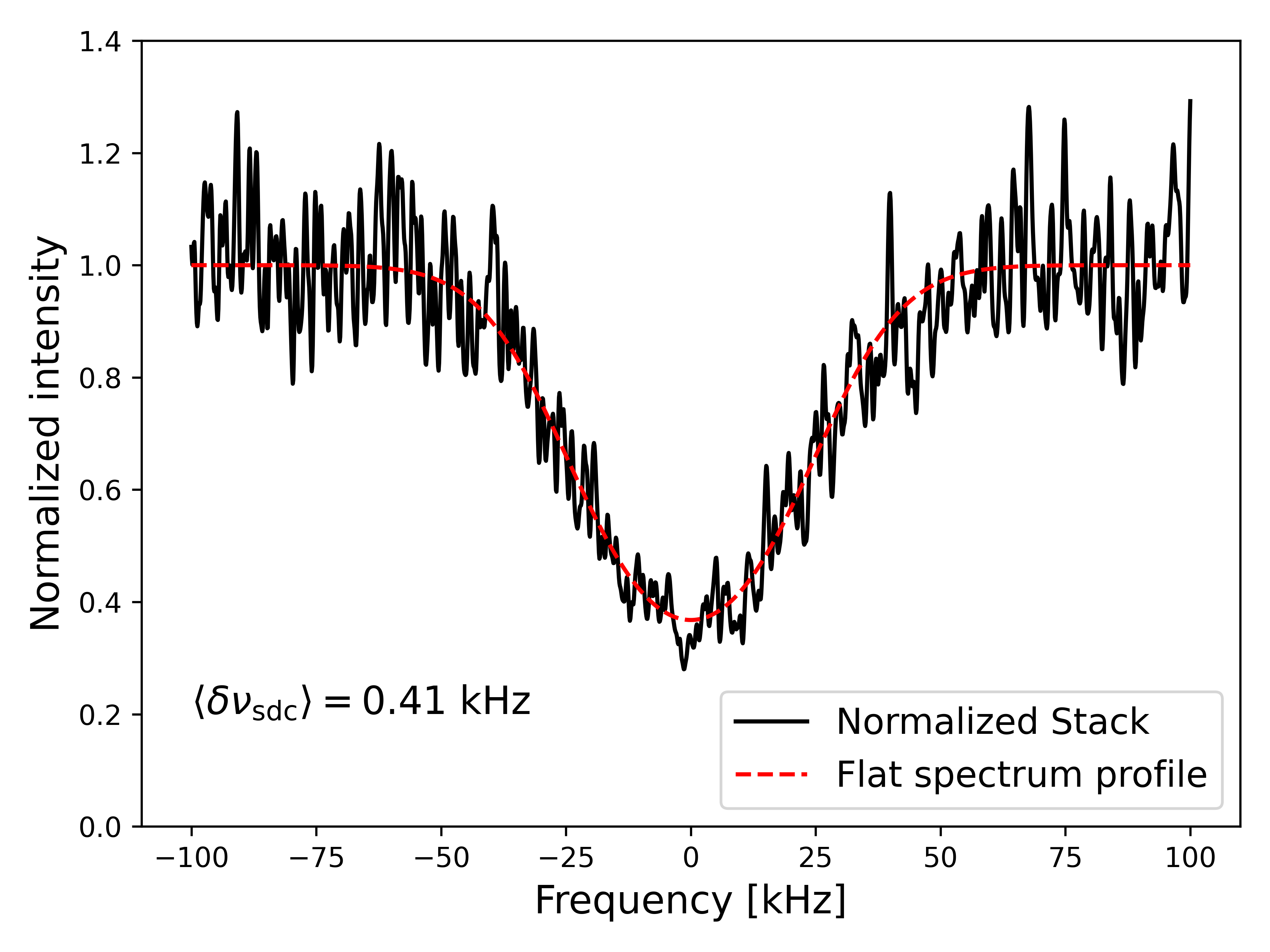}
    \caption{A stack of the spectra of 100 bursts from the same source, whose scintillation modulations are uncorrelated between bursts and for which the mean scintillation decorrelation bandwidth is $\langle \delta \nu_{\rm sdc} \rangle \approx 0.41$ kHz. The cloud properties are the same as those in Figure \ref{fig:full_abs_prof}, and $\delta \nu_{\rm obs} = 0.1$ kHz.}
    \label{fig:100stack}
\end{figure}

We recreate the stacking process with our simulation methodology described in Appendix \ref{sec:reduce_comp}, which does not make the assumptions required for the analytical result.
We plot a stacked spectrum of 100 uncorrelated bursts coming from the same hypothetical repeating source in Figure \ref{fig:100stack}. The intrinsic flux is assumed to be the same for all bursts. It is checked that the recovered $m_I\approx1$ for each burst and the average scintillation bandwidth for the sample $\langle \delta \nu_{\rm sdc} \rangle$ is close to the input value. The observed $m_{I, \rm off-line} = 0.0917$ compared to the theoretical expectation of $0.1$ for 100 bursts. The properties of the absorption line can be estimated by cross-correlating the spectrum with a grid of sample absorption templates and by including more than one Voigt/Gaussian profiles as required for multicomponent profiles. For the Milky Way, it can also benefit from prior knowledge of the cloud's velocity from CO or HI emission surveys, which would narrow the frequency range to search and improve detection significance.

\section{Observational Analysis}

\subsection{Milky Way Molecular Cloud Properties} \label{sec:molcloud}
A catalog of 8107 molecular clouds is investigated by \cite{Miville2017} covering the entire Galactic plane, including 98$\%$ of $^{12}$CO emission. They derive an average radius of molecular clouds $R\approx31.5$ pc. Considering the 13.5 kpc radius of the disk of the Milky Way, the covering fraction of these observed molecular clouds can be roughly calculated to be $f_{MC} \sim 0.044$. 

The filling factor of dense gas in these clouds as traced by CO emission is much less than unity \citep{Perault1985}, and in one case, has been found to be around 3\% \citep{Lizst2016}. Moreover, CO is used as a tracer for molecular hydrogen, which may occupy only the densest cores of molecular clouds. HI may occupy a larger fraction of the volume in molecular clouds than CO, although it is present outside the dense molecular regions. Translating to the size of HI envelopes of molecular clouds observed in CO is not possible without detailed abundance maps of each cloud.

\subsection{Analysis of a Cross-Reference Between FRB and Molecular Cloud Sightlines} \label{sec:cross-reference}
No galactic latitude dependence is expected for FRBs due to their extragalactic origin, and it is also not seen in data \citep{Josephy2021}. Consequently, a fraction of FRBs will lie along low galactic latitudes, and some of these FRB sightlines could intersect molecular clouds, which are primarily constrained to the plane of the Milky Way. To test this hypothesis, we match the sky positions of molecular clouds in the catalog of \citet{Miville2017} (see Section \ref{sec:molcloud} for a relevant discussion) with the sky positions of FRB repeaters in the Blinkverse database \citep{Blinkverse2023}. Nonrepeaters are not tested for cloud-aligned sightlines because instrumental sensitivities are not good enough to detect absorption profiles in a single burst. For example, a lack of detection is noted for all five low-latitude FRBs discovered by the Five-hundred-meter Aperture Spherical Telescope (FAST) GPPS survey (\citealt{Zhou2023}; by private communication with D. J. Zhou). Even for the extremely bright FRB 20180309, no indication of an absorption signal was detected \citep{Oslowski2019}. Similarly, FRB 20211127I has been recently tested for HI absorption without detection \citep{Roxburgh2026}.

FRB 20180916B (also known as R3; $\ell = 129.71^{\circ}$, $b=3.73^{\circ}$), is the only repeating FRB whose line of sight passes through (or close to) a molecular cloud. This cloud has a size $= 17.969 \,\rm pc$, is located at a kinematic distance of $\sim 3\,\rm kpc$, and has a mass $\sim 4\times10^4 M_{\odot}$. So far, only 61 bursts have been recorded from this well-studied source in L band \citep{Pastor-Marazuela2021, Bethapudi2025}. This source exhibits only Milky Way contributed scintillation, with $\delta \nu_{\rm dc}(1.7 \;{\rm GHz}) = 59\pm 13$ kHz \citep{Marcote2020} and $\delta \nu_{\rm dc}(600 \;{\rm MHz})=0.82\pm 0.06$ kHz \citep{Bethapudi2023}, both of which are mutually consistent and correspond to $\delta \nu_{\rm dc}(1.42 \;{\rm GHz}) \approx 25$ kHz, using the well-known Kolmogorov scaling exponent of 4.4. The observed resolution for the data in \citet{Pastor-Marazuela2021} is 195 kHz, which is not conducive for detection. We perform a case study for a potential detection from this source in Section \ref{sec:case_study_frb20180916b}.

\section{Inclusion of Detector Noise} \label{sec:detector_noise}

\subsection{Single Bursts}
\begin{figure*}
    \centering
    \includegraphics[width=0.48\linewidth]{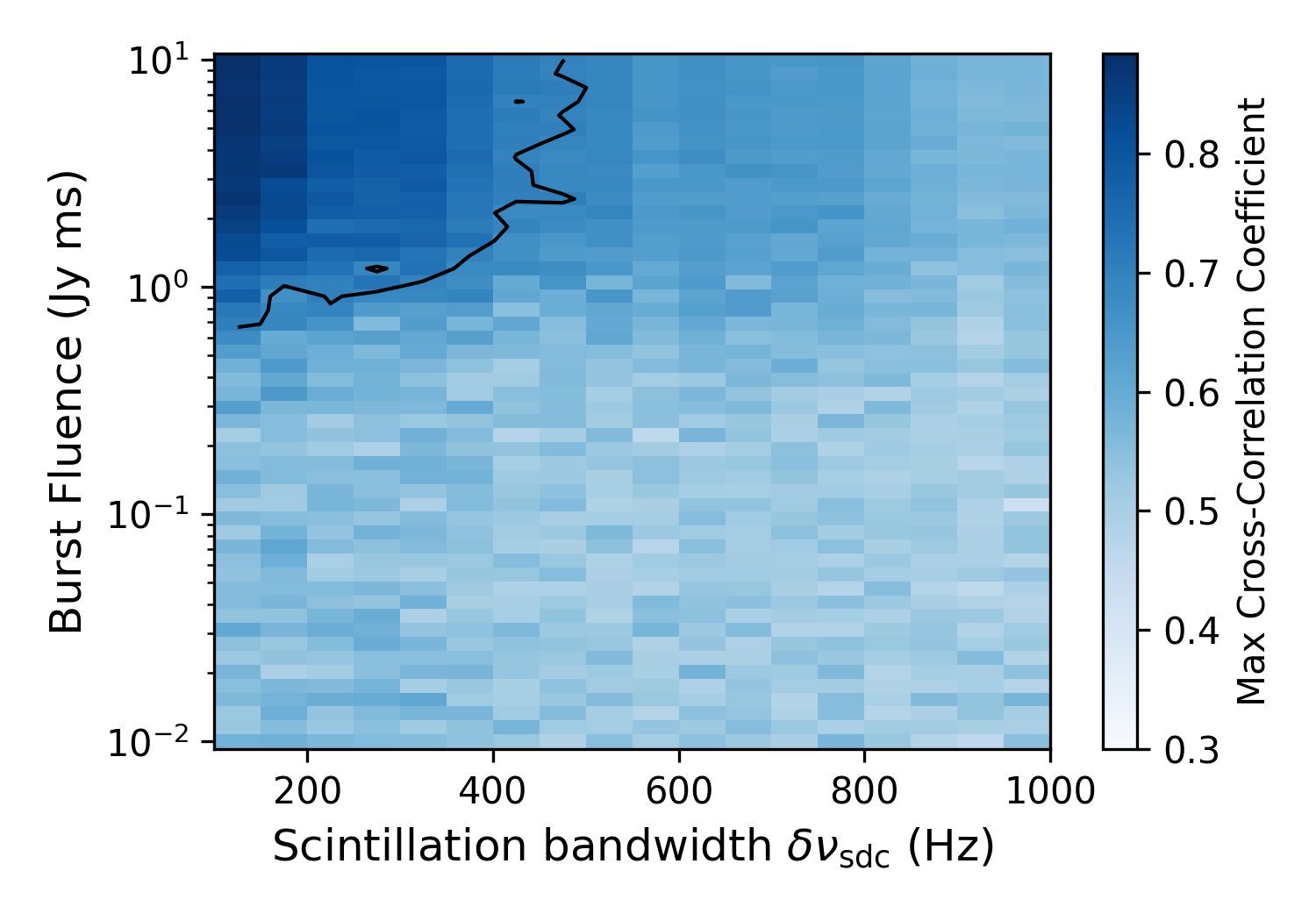}
    \hfill
    \includegraphics[width=0.48\linewidth]{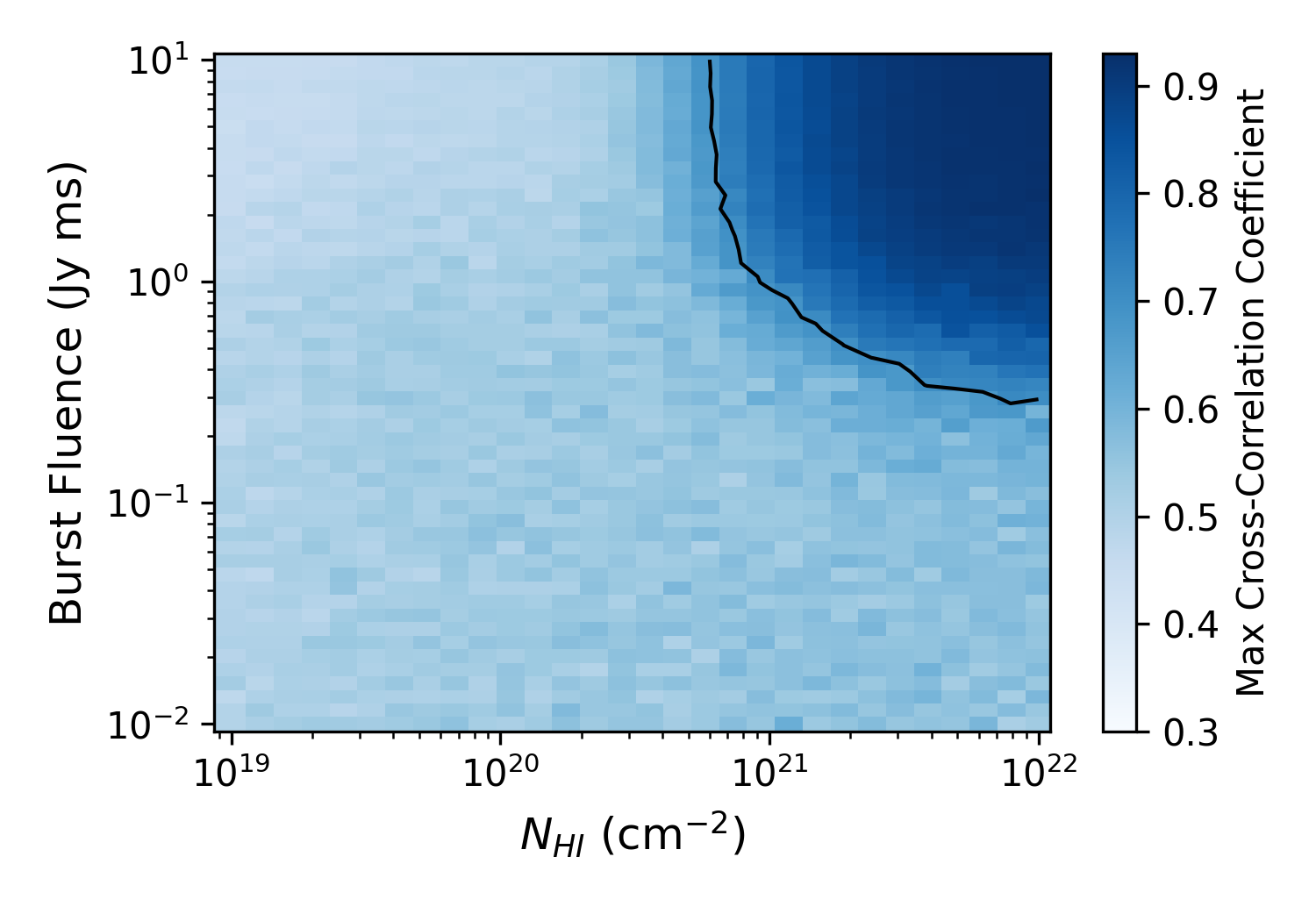}
    \caption{These plots show the averaged maximum cross-correlation coefficient (in color) obtained by cross-correlating the input flat spectrum profile (like the red dashed line in Figure \ref{fig:full_abs_prof}) with a single burst mock spectrum, scintillation and noise included, for FAST with SEFD $=1.36$ Jy. The vertical axis shows the single burst fluence measured at offline frequencies. ({\it left}) The horizontal axis shows the scintillation bandwidth, with the correlation coefficient being averaged over 20 bursts. $N_{\rm HI} \approx 9 \times 10^{20}$ cm$^{-2}$. ({\it right}) The horizontal axis shows the HI column density, with the correlation coefficient averaged over 50 bursts. $\langle \delta \nu_{\rm sdc} \rangle = 0.3$ kHz. For both plots, the black contour line indicates a maximum cross-correlation coefficient of 0.7. $d=20$ pc and $T_{\rm spin}=50$ K.}
    \label{fig:single_burst_param}
\end{figure*}

It is very important to note that up until now, detector noise has not been taken into account in this work. Note that when dealing with observations, we use flux and fluence interchangeably to relate theory to observations where the burst spectrum is obtained by integrating over time. Bright high signal-to-noise single bursts with a strong absorption line can yield its presence. When $\delta \nu_{dc}$ is much smaller than the FWHM of the absorption profile, we can run a low-pass filter on the signal that averages out the rapid scintillation and noise fluctuations. This can then be cross-correlated with template absorption profiles. To illustrate this in a simple way, we cross-correlate the exact input absorption spectrum with the output spectrum containing the absorption line, scintillation, and noise. The regime of scintillation bandwidths, HI column densities, and burst fluences where this cross-correlation yields a high confidence detection is highlighted in Figure \ref{fig:single_burst_param}. This figure is generated for FAST but can be scaled for any telescope with a known SEFD by scaling the vertical axis by a factor of SEFD$/1.36$. The methodology for introducing noise into the spectrum is discussed in Appendix \ref{app:detector_noise}.

Bright one-offs or somewhat rarer higher fluence bursts from repeaters in the Galactic plane are promising candidates. However, telescopes like FAST have a very low probability of detecting bright one-off FRBs due to their minuscule sky coverage \citep{Roxburgh2026}.
Furthermore, receiver (digital backend) saturation is a noted problem with FAST, where it detects only the lower-end of a repeater's fluence distribution compared to telescopes such as NRT \citep{Zhang2023, Konjin2024, Ould-Boukattine2026, Zhang2025}. This is why the fluence in Figure \ref{fig:single_burst_param} has been limited to a maximum value of 10 Jy ms.

\subsection{Repeat bursts} \label{sec:repeat_bursts}
\begin{figure*}
    \centering
    \includegraphics[width=0.49\linewidth]{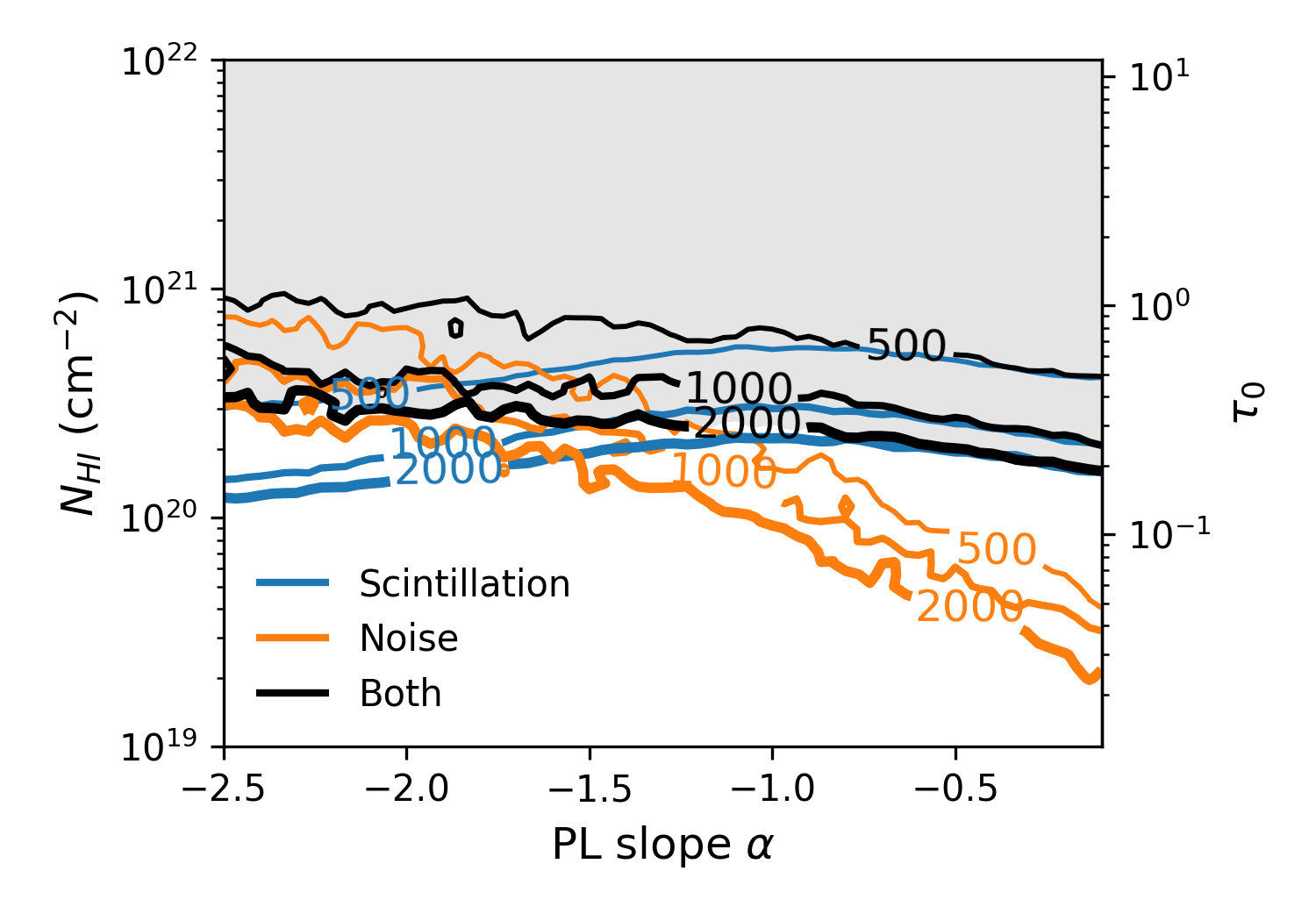}
    \hfill
    \includegraphics[width=0.49\linewidth]{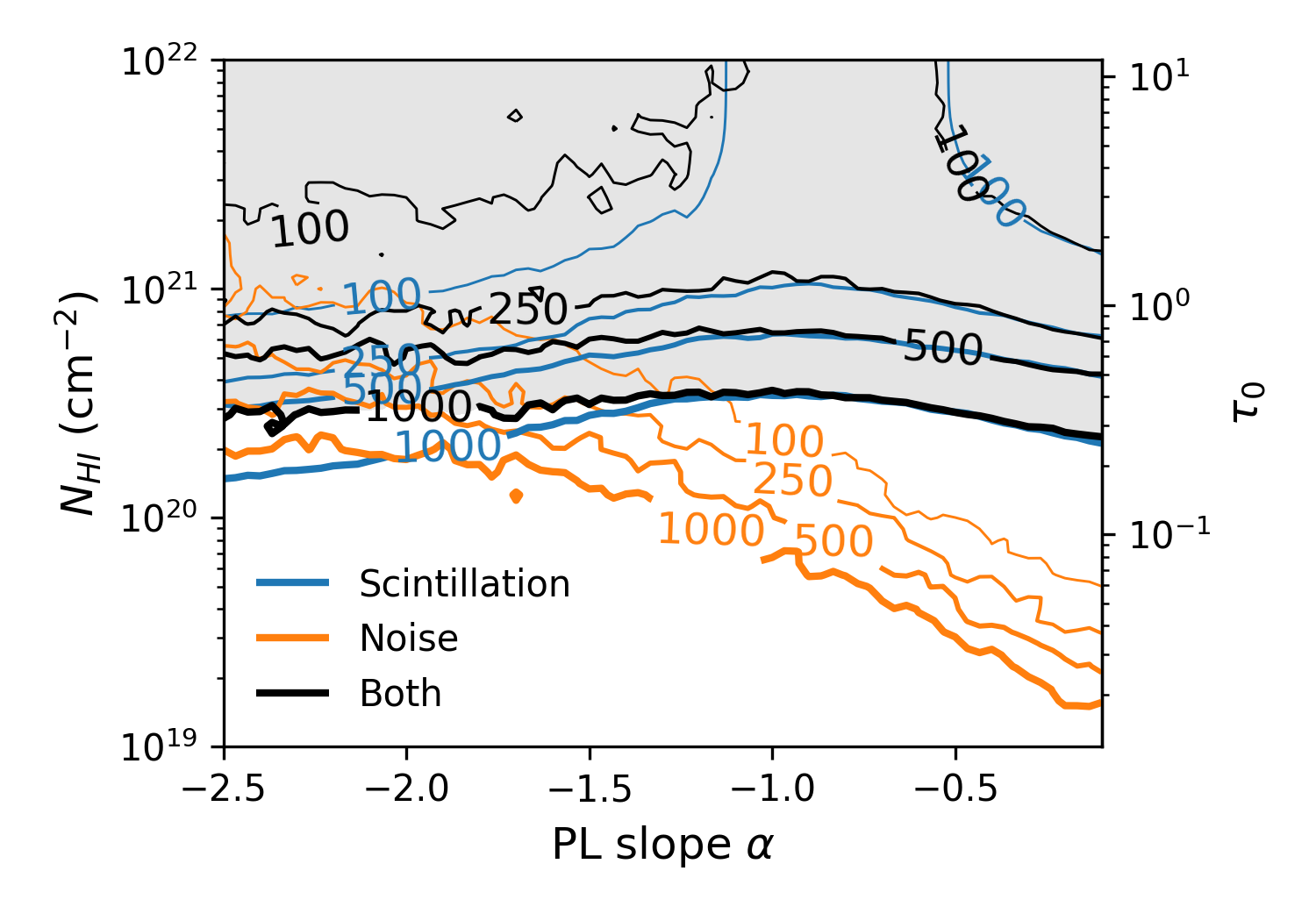}
    \caption{Shown are $5\sigma_{\rm stack}$ significance contours of HI absorption detection as a function of $N_{HI}$ and the power-law fluence CDF index $\alpha$. The contours highlight a comparison between cases when only scintillation is present (blue lines), when only detector noise is present (orange lines), and when both are present (black lines). ({\it left}) Mock observations correspond to FAST (SEFD $=1.36$ Jy) with stack sizes of 500, 1000, and 2000 bursts. The CDF is sampled between $0.05-5$ Jy ms. ({\it right}) Mock observations correspond to GBT (SEFD $=8.25$ Jy) with stack sizes of 100, 250, 500, and 1000 bursts. The CDF is sampled between $0.5-70$ Jy ms. $\langle \delta \nu_{\rm sdc} \rangle \approx 28$ kHz, and $\delta \nu_{\rm obs}=20$ kHz for both panels. The light gray shaded region shows the detectable parameter space with the corresponding panel's thickest contour. The right vertical axis shows the line-center optical depth ($\tau_0$; Equation \ref{eq:tau0}) based on the cloud properties, which are fixed to $T_{\rm spin} = 50$ K and $d\approx18$ pc.}
    \label{fig:scint_noise_compare}
\end{figure*}

Because the combination of cloud properties associated with a particular FRB and the detector noise is unique, it is difficult to estimate the number of bursts required for significant stacked detection a priori, for cases where $\delta \nu_{\rm dc} \sim $ FWHM. As an exercise, we consider cloud properties with $d=17.969$ pc (representative of the one found for FRB 20180916B) and $T_{\rm spin}=50$ K. We can define the standard deviation of flux fluctuations in the offline frequencies of a stacked spectrum as $\sigma_{\rm stack}$. Then, a map of the maximum depth of the absorption profile yields contour lines for a $5\sigma_{\rm stack}$ detection significance, which are shown in Figure \ref{fig:scint_noise_compare}. This figure has two panels, the left one for FAST and the right one for Green Bank Telescope (GBT) mock observations. Both panels are based on a general power-law CDF with no breaks, of the form $N(>F) = F^{\alpha}$, with $\alpha < 0$. The maximum depth values are the average of 100 iterations of sampling bursts from the fluence CDF. The panels highlight the regime where scintillation plays a more negative role than just detector noise. It is seen that the detection thresholds are roughly flat across the range of $\alpha$, except where a small number of bursts are being stacked. In such a case, it becomes impossible to detect the absorption line when $-1.0 \lesssim \alpha \lesssim -0.5$, as is seen for the 100 burst-stack contour in the right panel of Figure \ref{fig:scint_noise_compare}. Typically, repeaters have flatter observed CDF slopes at the low fluences probed by FAST, and steeper slopes at higher fluences \citep{Ould-Boukattine2026, Zhang2023}. FAST is currently the most sensitive detector with a system-equivalent flux density (SEFD) of 1.36 Jy at 1.42 GHz \citep{Jiang2020}, while GBT is the most sensitive detector that can actually observe FRB 20180916B. Baseband detections are preferred because they allow the frequency resolution to be optimized.

\subsection{Case study of FRB 20180916B} \label{sec:case_study_frb20180916b}
\begin{figure}
    \centering
    \includegraphics[width=\linewidth]{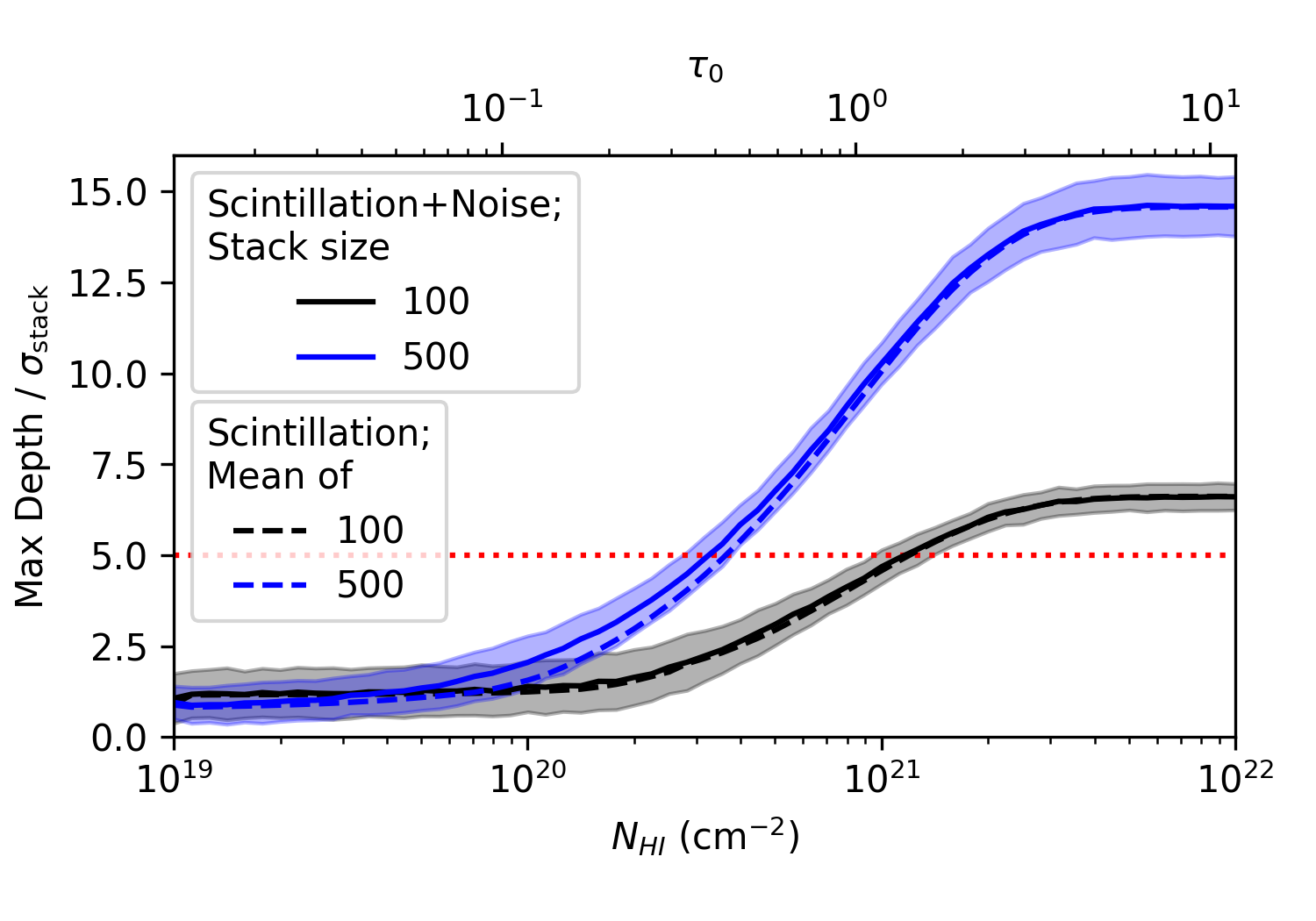}
    \caption{Significance of the maximum HI absorption depth achieved as a function of the HI column density of a single cloud ($N_{HI}$). The significance is given in terms of the standard deviation of the stacked flux fluctuations in the offline frequencies ($\sigma_{\rm stack}$). The dotted red line marks the $5 \sigma_{\rm stack}$ level. The black and blue solid lines and shaded regions mark the significance in GBT L-band mock observations of FRB 20180916B with a broken power-law fluence distribution, for stacks of 100 and 500 bursts, respectively.
    Dashed lines correspond to scintillation-only no-noise stacks.
    The mean scintillation decorrelation bandwidth $\langle \delta \nu_{\rm sdc} \rangle \approx 28$ kHz, akin to FRB 20180916B at 1.42 GHz, and $\delta \nu_{\rm obs}=20$ kHz. $d\approx 18$ pc and $T_{\rm spin}=50$ K.
    }
    \label{fig:1d_NHI_frb20180916b}
\end{figure}

FRB 20180916B has a broken power-law fluence CDF with index $\alpha=-1.4$ for $F>7.8$ Jy ms and $\alpha=-0.7$ for $F < 7.8$ Jy ms \citep{Pastor-Marazuela2021}. The break at $F=3.2$ Jy ms is associated with Westerbork/Aperitif sensitivity and is ignored. Thus, a power-law CDF with a single break at $F=7.8$ Jy ms, a maximum fluence of 70 Jy ms, and a minimum fluence of 0.5 Jy ms can be assumed for mock GBT observations. The minimum fluence is chosen by noting a turnover in the fluence distribution at 0.5 Jy ms for FRB 20240114A with GBT \citep{Xie2025}, which could either be intrinsic to the repeater or related to GBT's sensitivity. A test of the significance of detection, done via the maximum absorption line depth (mean of 100 fluence CDF sampling iterations) reported as a multiple of $\sigma_{\rm stack}$, is shown in Figure \ref{fig:1d_NHI_frb20180916b}. The dashed lines, representing scintillation-only runs, are close to the solid lines, which represent mock observations, showing that scintillation dominates the observational hurdles for FRB 20180916B.

\section{Discussion and Conclusions} \label{sec:conclusion}
This work studies the effects of scintillation on the detectability of 21 cm HI absorption and provides a clear pathway to HI absorption line detection. The ability to detect absorption lines with FRBs would provide a unique window into probing molecular clouds in great detail, with sightlines probing transverse scales $\sim$astronomical units. This is akin to probing with pulsars, for clouds within the Milky Way but is the only way to probe HI clouds with the same precision in host/intervening galaxies. Additionally, it would help probe cold HI gas in minihalos and other localized pockets of cold gas at high redshift during the Epoch of Reionization.

We present a highly efficient model of simulating scintillation arising in a thin scattering screen, which includes the effects of both geometric phase shift and the phase shift of a wave traveling through turbulent plasma. We show that flux modulations due to scintillation, which can hide the absorption line, can be overcome if the scintillation bandwidth is quite different from the width of the absorption line. Specifically, if $\delta \nu_{\rm dc} \ll $ FWHM of the HI line, a low-pass filter may be used to average out the flux fluctuations. Such a low-pass filter may simply be the observing resolution $\delta \nu_{\rm obs}$, as long as $\delta \nu_{\rm dc} \ll \delta \nu_{\rm obs} <$ FWHM. Using this method, we identify the range of scintillation and cloud properties where bright single bursts can yield the absorption line. While a sensitive telescope like FAST is disadvantaged at detecting one-off bursts, upcoming telescopes, like CHORD \citep{Vanderlinder2019}, will be able to do so with good sensitivity. It is noted that scintillation might not be a concern at high Galactic latitudes, unless the host galaxy has scintillation bandwidth of order the absorption linewidth. Galactic scintillation could also be of concern for high latitude sightlines, when absorption lines of high-redshift FRBs are shifted to a few $\times$ 100 MHz in the observer frame.

The deleterious effect of scintillation can also be overcome if a sufficient number of repeating FRB spectra are stacked to cancel out the flux modulations. Figure \ref{fig:scint_noise_compare} shows the parameter space where burst stacks can lead to a significant absorption detection, with both FAST and GBT yielding similar detection contours in their respective fluence ranges. We have shown that for flatter fluence distribution slopes, the HI absorption line detectability is more scintillation-limited than noise-limited. This goes beyond previous works, which have just considered detector noise, such as \citet{Roxburgh2026}. Apart from stacking repeaters, bursts from redshift-identified nonrepeaters can be used to stack their source-frame spectra. In this case, different scintillation frequency scales are imprinted on the stacked spectrum, but with a deepening absorption profile at and around 1.42 GHz in the source frame.

We find from an analysis of Milky Way molecular clouds, that the repeating FRB 20180916B's sightline might intersect a Galactic molecular cloud. Such a repeater is valuable in unlocking FRBs as tools to probe neutral hydrogen. Figure \ref{fig:1d_NHI_frb20180916b} shows that only 100 bursts with GBT may be enough to detect an absorption line arising from a cloud $N_{HI} \gtrsim 10^{21}$ cm$^{-2}$.

There is sufficient detection capability to perform HI absorption studies from sources at low redshifts. Still, improvements in telescope sensitivities are important to enable absorption line detection in the steeper slope regime of the fluence distribution. Furthermore, improved sensitivities will allow the detection horizon to increase in redshift. This will increase the scope of HI detection from galaxies to high-redshift structures, such as minihalos. Even though \citet{Gupta2025} find that the detectability of $z>6$ FRBs will remain a challenge for planned telescopes, HI absorption profile detection in high-redshift FRBs may be a viable long-term possibility.

\begin{acknowledgments}
    Code for exploring and producing the data is provided in the Github repository: \url{https://github.com/omguptaup/frb-h1absorb}.
    
    The authors thank the referee for constructive criticisms that significantly improved the manuscript. O.G. would like to thank Danielle Berg for the course on the Interstellar Medium, which inspired this project. O.G. also thanks Dejiang Zhou, Kaustubh Rajwade, Sachin Pradeep, Neal Evans, Alex Cooper, and Kenzie Nimmo for helpful discussions and/or comments. O.G. also acknowledges fruitful discussions with several attendees of the FRB 2023 conference where an earlier version of this work was first presented.

    The authors acknowledge the Texas Advanced Computing Center (TACC) at The University of Texas at Austin for providing HPC resources through project AST22018. O.G. acknowledges the use of GitHub Copilot code completion on Microsoft Visual Studio Code, and the use of Google Gemini as a search tool. The authors also acknowledge that UT Austin, where this research has been conducted, is on the Indigenous land of the Tonkawa people, and historically, the Comanche and Apache moved through this area.

    OG has been supported by the J. Craig Wheeler Endowed Fellowship in Astronomy (UT Austin) for a part of the project's duration. PK's work was funded in part by an NSF grant AST-2009619. P.K., P.B., and O.G. acknowledge support provided by the NASA grant 80NSSC24K0770 for this work. PB's work was also funded by a grant (no. 2024788) from the United States-Israel Binational Science Foundation (BSF), Jerusalem, Israel and by a grant (no. 1649/23) from the Israel Science Foundation.
\end{acknowledgments}

\appendix

\section{Model for Galactic Distribution of Molecular Clouds}
\label{app:galdist_model}
We generate a simple model for the distribution of molecular clouds in Milky Way-like galaxies. The radial profile is assumed to be exponentially decreasing, and the vertical distribution is taken to be a Gaussian function with a uniform scale height \citep{Jeffreson2022}. For disks with galactic radius $R_{\rm gal} < 5$ kpc, the scale height $H(R_{\rm gal})$ is taken to be linearly increasing with $R_{\rm gal}$, from 35 pc to 50 pc, whereas, $H(R_{\rm gal} > 5 \;{\rm kpc}) = 50 $ pc. Then, the number density of clouds can be defined as,
\begin{equation}
    n_c = n_0 \exp \Big[ \frac{-2r}{R_{\rm gal}} \Big] \exp\Big[ \frac{-z^2}{2 H(R_{\rm gal})^2} \Big].
\end{equation}
The value of $n_0$ is obtained by setting $N_c = \int n_c dV$ to the total number of molecular clouds. For the Milky Way, $N_c \approx 8000$ \citep{Miville2017}, and the volume integral takes $R_{\rm gal} = 13.5$ kpc and the vertical height of the Galactic disk to be 180 pc.

\begin{figure}
    \centering
    \includegraphics[width=0.8\linewidth]{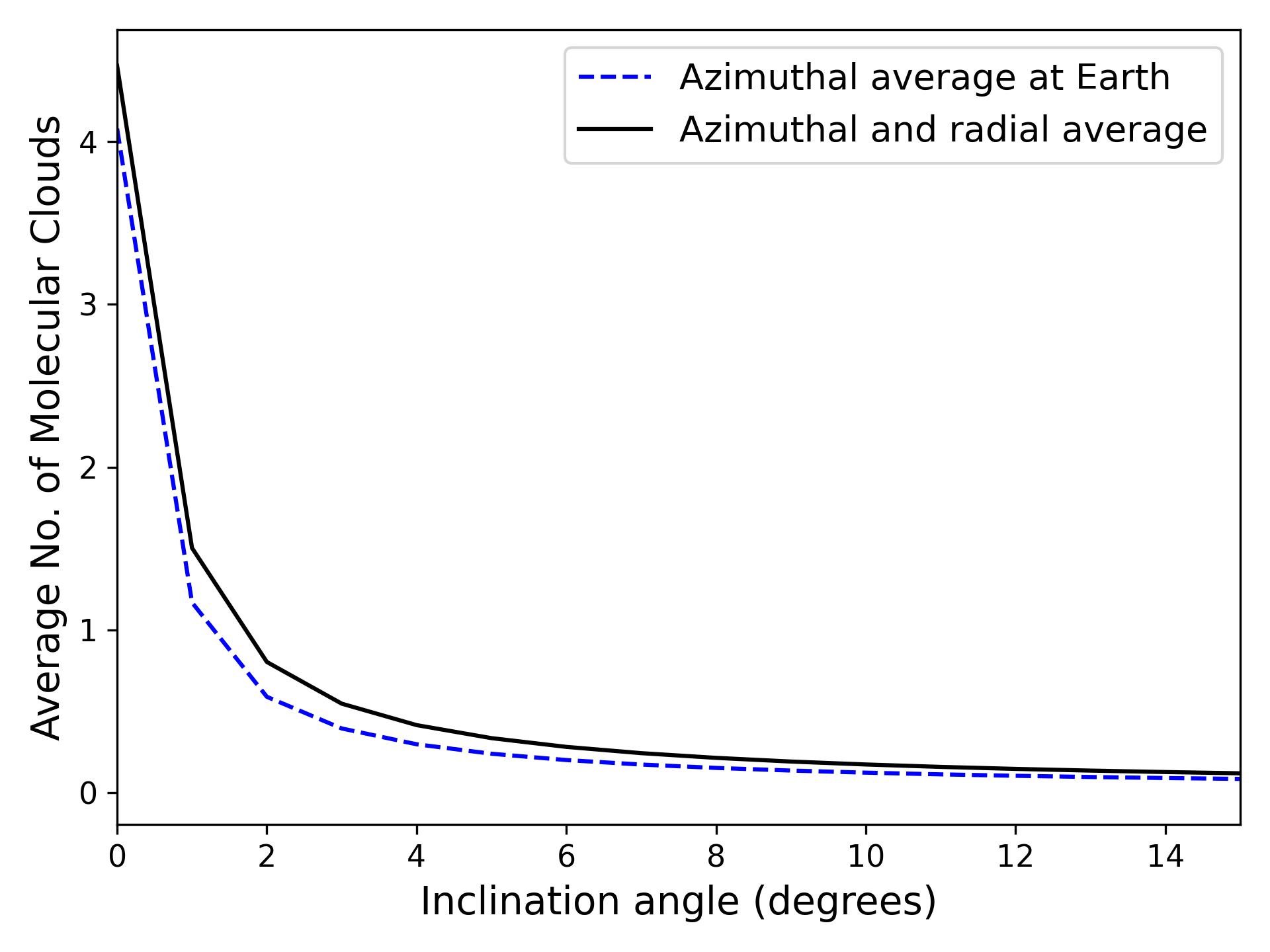}
    \caption{Average number of molecular clouds intersected by sightlines at different inclination angles from the Galactic plane. The dashed blue curve shows the average over all azimuthal angles for a point 8 kpc from the Galactic center (roughly the Sun/Earth distance), while the solid black line corresponds to the average over all azimuthal angles as well as over all points along the radial direction.}
    \label{fig:avgcloudnum_inc}
\end{figure}

Using this model, simple estimates of the average number of molecular cloud intersections can be made, to establish that low-latitude sightlines are more likely to encounter molecular clouds. Figure \ref{fig:avgcloudnum_inc} shows the average number of molecular cloud intersections as a function of latitude, which increases to a value greater than 1 at latitudes lower than 2 degrees. The fraction of sightlines from a point 8 kpc from the Galactic center (equivalent to the solar position), which encounter at least 1 molecular cloud, is 1.78\%. This is a small value because not only low-latitude sightlines, but additionally those toward the Galactic center, are preferred. For example, a Galactic center sightline from a point 8 kpc from the center in the plane of the disk will encounter $\approx 11$ clouds, whereas a sightline toward the anticenter will encounter roughly 1 cloud.

\section{Further reduction of the computational complexity in Simulating the scintillation bandwidth} \label{sec:reduce_comp}
Simulating a scattering screen is a very computationally intense process when the scintillation bandwidth $\lesssim 100$ kHz. This is because a large screen area (i.e., large number of eddies) needs to be simulated at 1.42 GHz (see Equation \ref{eq:screen_size}). However, the problem can be alleviated by implementing certain steps. First, from Equation (\ref{eq:decorr_bandwidth}), it is noted that $\delta \nu_{\rm dc} \propto \nu^{4.4}$, and as such, the simulation can be run at a significantly lower frequency than $1.42$ GHz, provided the central simulation frequency is much larger than the total bandwidth being simulated. Second, the frequency dependence for $\phi_{\rm s}$ is retained, which gives rise to the small $\delta \nu_{\rm dc}$ required. However, $\delta \nu_{\rm dc}$ changes rapidly with frequency for small central frequencies, as is chosen here. The choice of the central frequency being much larger than the bandwidth being simulated helps in this regard, as $\delta \nu_{\rm dc}$ changing over the bandwidth is not apparent. Since, the scintillation should have flux modulations consistent with the central frequency of 1.42 GHz, where scintillation properties do not change over small frequency intervals of upto a few MHz, we explicitly remove the frequency dependence of $R_F$. This leads to a frequency-independent $\phi_{\rm g}$, which makes the corresponding phase contribution to $\psi$ in Equation (\ref{eq:fresnel_kirchoff_integral}) and the denominator constant with frequency.

\section{Quantifying Detector Noise}
\label{app:detector_noise}

The noise in the detector is given by the radiometer equation,
\begin{equation}
    \sigma = \frac{T_{\rm sys}}{G \sqrt{n_{\rm pol} \delta\nu_{\rm obs} \Delta t}},
\end{equation}
where, $T_{\rm sys}$ is the nominal system temperature when the source is off, $G$ is the system gain in K/Jy, $n_{\rm pol}$ is the number of polarizations used for observations, $\delta \nu_{\rm obs}$ is the observational frequency resolution, and $\Delta t$ is the time over which the spectrum is integrated.

If $S_{\nu, sc}$ is the simulated scintillation flux in each frequency bin of width $\delta \nu_{\rm obs}$, then $S_{\rm \nu, tot} = S_{\nu, sc} + \mathcal{N}(0, \sigma)$, where, $\mathcal{N}$ represents the normal distribution with mean and standard deviation given in the parentheses. 
When stacking $N$ bursts, and assuming that each burst lasts for a duration $\delta t$, the spectrum integration time $\Delta t \approx N \delta t$. Then, the noise $\sigma \propto N^{-1/2}$. 


\bibliography{sample7}{}
\bibliographystyle{aasjournalv7}



\end{document}